# Confidence intervals of prediction accuracy measures for multivariable prediction models based on the bootstrap-based optimism correction methods

Hisashi Noma, PhD*
Department of Data Science, The Institute of Statistical Mathematics, Tokyo, Japan
ORCID: http://orcid.org/0000-0002-2520-9949

Tomohiro Shinozaki, PhD
Department of Information and Computer Technology, Faculty of Engineering, Tokyo University of Science, Tokyo, Japan

Katsuhiro Iba
Department of Statistical Science, School of Multidisciplinary Sciences, The Graduate University for Advanced Studies, Tokyo, Japan
Office of Biostatistics, Department of Biometrics, Headquarters of Clinical Development, Otsuka Pharmaceutical Co., Ltd., Tokyo, Japan

Satoshi Teramukai, PhD
Department of Biostatistics, Graduate School of Medical Science, Kyoto Prefectural University of Medicine, Kyoto, Japan

Toshi A. Furukawa, MD, PhD
Departments of Health Promotion and Human Behavior and of Clinical Epidemiology, Kyoto University Graduate School of Medicine/School of Public Health, Kyoto, Japan

*Corresponding author: Hisashi Noma
 Department of Data Science, The Institute of Statistical Mathematics
 10-3 Midori-cho, Tachikawa, Tokyo 190-8562, Japan
 TEL: +81-50-5533-8440
 e-mail: noma@ism.ac.jp

**Abstract**

In assessing prediction accuracy of multivariable prediction models, optimism corrections are essential for preventing biased results. However, in most published papers of clinical prediction models, the point estimates of the prediction accuracy measures are corrected by adequate bootstrap-based correction methods, but their confidence intervals are not corrected, e.g., the DeLong's confidence interval is usually used for assessing the *C*-statistic. These naïve methods do not adjust for the optimism bias and do not account for statistical variability in the estimation of parameters in the prediction models. Therefore, their coverage probabilities of the true value of the prediction accuracy measure can be seriously below the nominal level (e.g., 95%). In this article, we provide two generic bootstrap methods, namely (1) location-shifted bootstrap confidence intervals and (2) two-stage bootstrap confidence intervals, that can be generally applied to the bootstrap-based optimism correction methods, i.e., the Harrell's bias correction, 0.632, and 0.632+ methods. In addition, they can be widely applied to various methods for prediction model development involving modern shrinkage methods such as the ridge and lasso regressions. Through numerical evaluations by simulations, the proposed confidence intervals showed favourable coverage performances. Besides, the current standard practices based on the optimism-uncorrected methods showed serious undercoverage properties. To avoid erroneous results, the optimism-uncorrected confidence intervals should not be used in practice, and the adjusted methods are recommended instead. We also developed the R package `predboot` for implementing these methods (https://github.com/nomahi/predboot). The effectiveness of the proposed methods are illustrated via applications to the GUSTO-I clinical trial.

Key words: multivariable prediction model; discrimination and calibration; optimism; bootstrap; confidence interval.

# 1. Introduction

In the development of clinical prediction models, multivariable prediction models have been essential statistical tools for incorporating multiple predictive factors to construct diagnostic and prognostic algorithms [1,2]. A multivariable prediction model is usually constructed by an adequate regression model (e.g., a logistic regression model for a binary outcome) based on a series of representative patients from the source population, but their "apparent" predictive performances such as discrimination and calibration measures are biased from their actual performances for external populations [3,4]. This bias is known as "optimism" in prediction models. Practical guidelines (e.g., the TRIPOD statements [3,4]) recommend adopting principled optimism adjustment methods for internal validations, and these are currently the standard statistical analysis methods in practice. In particular, the bootstrap-based correction methods, i.e., the Harrell's bootstrapping bias correction [1], 0.632 [5], and 0.632+ [6] methods, have been recommended [2,7].

However, the optimism corrections are mainly applied only to point estimates of the prediction performance measures in current practice. Even in recent leading medical journals, although many papers provided optimism-corrected estimates, their confidence intervals were usually not optimism-corrected, e.g., for *C*-statistics, and many papers provided solely the conventional DeLong's confidence interval [8]. Since the point estimate of the naïve, uncorrected predictive accuracy measures is biased, the actual coverage rates of their confidence intervals can be seriously below the nominal level (e.g., 95%). The reported predictive performance estimates can directly influence clinical guidelines or medical practices, so assuring the validity of their inferences is a relevant problem.

In previous methodological studies, resampling-based confidence intervals have been discussed for CV [9,10]. However, there have been no effective techniques to construct valid confidence intervals based on the bootstrap-based optimism correction methods. In this article, we propose effective methods to construct the confidence intervals, particularly we provide two generic bootstrap algorithms, namely (1) location-shifted bootstrap confidence intervals and (2) two-stage bootstrap confidence intervals, that can be widely applied to various approaches to prediction model development involving modern shrinkage methods such as the ridge [11] and lasso [12] regressions. In numerical evaluations using simulations, the proposed confidence intervals based on the three optimism correction methods showed favourable coverage performances. However, the current standard practices based on the optimism-uncorrected methods showed marked undercoverage properties. We also illustrate the effectiveness of the proposed methods via applications to real-world data from the GUSTO-I trial [13,14].

## 2. Estimation of prediction accuracy measures

### 2.1 Logistic regression model for clinical predictions

First, we briefly introduce the fundamental methods for multivariable prediction models and their prediction accuracy measures. In this article, we consider to construct a logistic regression prediction model for a binary outcome [15], but the proposed methods can similarly be applied to other types of prediction models, e.g., the Cox regression for censored time-to-event outcomes [1,16]. We denote $y_i$ $(i = 1,2,...,n)$ as a binary outcome variable ( $=1$ : event occurrence, or $=0$: not occurrence) and $\boldsymbol{x}_i = (x_{i1}, x_{i2}, ..., x_{ip})^T$ $(i = 1,2,...,n)$ as $p$ predictor variables for $i$th individual. The probability of event occurrence $\pi_i = \Pr(y_i = 1|\boldsymbol{x}_i)$ is modelled by the logistic regression model as

$$\pi_i = \frac{\exp(\beta_0 + \beta_1 x_{i1} + \beta_2 x_{i2} + \cdots + \beta_p x_{ip})}{1 + \exp(\beta_0 + \beta_1 x_{i1} + \beta_2 x_{i2} + \cdots + \beta_p x_{ip})}$$

where $\boldsymbol{\beta} = (\beta_0, \beta_1, ..., \beta_p)^T$ is the regression coefficient vector. Plugging an appropriate estimate $\widehat{\boldsymbol{\beta}}$ into the above model for $\boldsymbol{\beta}$, the risk score $\hat{\pi}_i$ $(i = 1,2,...,n)$ is defined as the estimated probability of individual patients. This risk score is used as the criterion to determine the predicted outcome [2,15].

For estimating the regression coefficients $\boldsymbol{\beta}$, the most popular conventional approach is the maximum likelihood (ML) estimation. The ordinary ML estimation can be easily implemented by standard statistical packages and has favourable theoretical properties such as asymptotic efficiency [17]. However, the ML-based modelling strategy is known to have several finite sample problems, e.g., when applied to a small or sparse dataset [18-21]. To address these problems, several alternative effective estimation methods have been developed. Representative approaches are the shrinkage regression methods such as the ridge [11,22] and lasso [12] regressions that use penalized likelihood functions to estimate the regression coefficients. Through the penalizations, the resultant regression coefficient estimates are shrunk towards zero and thereby can reduce overfitting. Also, a number of these estimating methods can shrink some regression coefficients to be exactly 0 via the functional types of penalty terms (e.g., lasso [12]), and can simultaneously perform variable selection. Other prediction algorithms involving machine learning methods have been well investigated [2,23], and the following proposed methods are generally applied to these methods.

*2.2 Prediction accuracy measures*

For assessing the predictive performances of the developed multivariable models, several accuracy measures are considered regarding their discrimination and calibration abilities [2]. Discrimination refers to the ability to classify high- and low-risk patients, and the most commonly used measure is the *C*-statistic, which assesses the concordance of the predicted and observed outcomes [1]. The *C*-statistic also corresponds to the area under the curve (AUC) of the empirical receiver operating characteristic (ROC) curve for the risk score [2]. The *C*-statistic ranges from 0.5 to 1.0, with larger values corresponding to superior discriminant performance. Calibration refers to the ability to determine whether the predicted probabilities agree with the observed probabilities [2]. The calibration plot [24] is widely used for assessing the concordance between the predicted and observed probabilities, and well-calibrated models have a slope of 1 for the linear regression of these two quantities. The proposed methods in Section 3 can generally be applied to these predictive measures.

All of these prediction accuracy measures may have biases from their actual performances for external populations if they are assessed for the derivation dataset itself [3,4]. This bias is known as optimism in prediction models. To assess these accuracy measures appropriately, adequate optimism adjustments are needed, and practical guidelines (e.g., the TRIPOD statements [3,4]) recommend using principled internal validation methods, e.g., split-sample, CV, and bootstrap-based corrections. Among these validation methods, split-sample analysis is known to provide a relatively imprecise estimate, and CV is not suitable for some performance measures [3,4,7]. Thus, bootstrap-based methods are generally recommended [3,4,7]. In Sections 2.3-2.5, we briefly review the three bootstrap-based methods, namely the Harrell's bias correction, 0.632, and 0.632+ methods.

*2.3 Harrell's bias correction*

Currently, the most widely applied bootstrap-based correction method in practice is Harrell's bias correction [1], which is based on the conventional bootstrap bias correction [25,26]. The algorithm is summarized as follows:

1. Let $\hat{\theta}_{app}$ be the apparent estimate for predictive accuracy measure of the original

population.

2. Conduct $B$ bootstrap resamplings with replacement from the original population.

3. Build prediction models for the $B$ bootstrap samples, and compute the predictive accuracy measure estimates for them, $\hat{\theta}_{1,boot}, \hat{\theta}_{2,boot}, \cdots, \hat{\theta}_{B,boot}$.

4. By the prediction models constructed from the $B$ bootstrap samples, compute the predictive accuracy measure estimates for the original population, $\hat{\theta}_{1,orig}, \hat{\theta}_{2,orig}, \cdots, \hat{\theta}_{B,orig}$.

5. The optimism estimate is provided as

$$\hat{\Lambda} = \frac{1}{B}\sum_{b=1}^{B}(\hat{\theta}_{b,boot} - \hat{\theta}_{b,orig})$$

The bias-corrected estimate is obtained by subtracting the estimate of optimism from the apparent performance, $\hat{\theta}_{app} - \hat{\Lambda}$.

The bias-corrected estimate is calculable by a relatively simple algorithm, and some simulation-based numerical evidence has shown that it has favourable properties under realistic situations [7,27]. However, a certain proportion of patients in the original population (approximately 63.2%, on average) should be overlapped in the bootstrap sample. The overlap may cause overestimation of the predictive performance [23], and several alternative methods have therefore been proposed.

*2.4 The 0.632 method*

The 0.632 method [5] was proposed as another bias correction technique to address the overlapping problem. In each bootstrapping, we formally regard the "external" subjects who are not included in the bootstrap sample as a "test" dataset for the prediction model developed in the bootstrap sample. Then, we compute the estimates of predictive accuracy measure for the external samples by the $B$ prediction models $\hat{\theta}_{1,out}, \hat{\theta}_{2,out}, \cdots, \hat{\theta}_{B,out}$, and denote the mean as $\hat{\theta}_{out} = \sum_{b=1}^{B} \hat{\theta}_{b,out}/B$. Thereafter, the 0.632 estimator is defined as a weighted average of the predictive accuracy measure estimate in the original sample $\hat{\theta}_{app}$ and the external sample estimate $\hat{\theta}_{out}$:

$$\hat{\theta}_{0.632} = 0.368 \times \hat{\theta}_{app} + 0.632 \times \hat{\theta}_{out}$$

The weight 0.632 is derived from the approximate proportion of subjects included in a bootstrap sample. Since the subjects that are included in a bootstrap sample are independent from those that are not, the 0.632 estimator can be interpreted as an extension of CV. However, the 0.632 estimator is associated with overestimation bias under highly overfit situations, when the apparent estimator $\hat{\theta}_{app}$ has a large bias [6].

### 2.5 The 0.632+ method

Efron and Tibshirani [6] proposed the 0.632+ method to address the problem of the 0.632 estimator. They introduced a relative overfitting rate $R$ as

$$R = \frac{\hat{\theta}_{out} - \hat{\theta}_{app}}{\gamma - \hat{\theta}_{app}}$$

$\gamma$ corresponds to "no information performance", which is the predictive performance measure for the original population when the outcomes are randomly permuted. The overfitting rate $R$ approaches 0 when there is no overfitting ($\hat{\theta}_{out} = \hat{\theta}_{app}$), and approaches 1 when the degree of overfitting is large. Then, the 0.632+ estimator is defined as

$$\hat{\theta}_{0.632+} = (1 - w) \times \hat{\theta}_{app} + w \times \hat{\theta}_{out}$$

$$w = \frac{0.632}{1 - 0.368 \times R}$$

The weight $w$ ranges from 0.632 ($R = 0$) to 1 ($R = 1$). Hence, the 0.632+ estimator gets closer to the 0.632 estimator when there is little overfitting, and gets closer to the external sample estimate $\hat{\theta}_{out}$ when there is marked overfitting.

## 3. Confidence intervals of the prediction accuracy measures

### 3.1 Location-shifted bootstrap confidence interval

The bootstrap-based optimism correction methods are essential for bias corrections of the prediction measures, but currently there are no effective methods based on the bootstrap-based corrections to construct confidence intervals that consider the optimism. Conventional analytical approaches for apparent measures (e.g., the DeLong's confidence interval for *C*-statistic [8]) and the naïve bootstrap confidence interval should provide invalid confidence intervals under realistic situations due to the biases.

The first approach we propose here is the location-shifted bootstrap confidence interval. This approach is simple. Based on the asymptotic theory for the bootstrap [25,26], the naïve bootstrap confidence interval for the apparent measures can adequately evaluate the approximate statistical variability of the predictive measures under large sample settings. However, its location should be shifted upwardly by the bias of the point estimate. Thus, the invalidity of the naïve bootstrap confidence interval is expected to be addressed if the bias of the location is adjusted when the sample size is sufficiently large. The proposal here is to correct the "location" of the naïve bootstrap confidence interval by the estimated bias. The algorithm to calculate the confidence limits is provided as follows.

*Algorithm 1 (Location-shifted bootstrap confidence interval)*

1. For a multivariable prediction model, let $\hat{\theta}_{app}$ be the apparent predictive measure for the derivation population and let $\hat{\theta}$ be the optimism-corrected predictive measure obtained from the Harrell's bias correction, 0.632, or 0.632+ method.

2. In the computational processes of $\hat{\theta}$, we can obtain a bootstrap estimate of the sampling distribution of $\hat{\theta}_{app}$ from the $B$ bootstrap samples. Compute the bootstrap confidence interval of $\hat{\theta}_{app}$ from the bootstrap distribution, $(\hat{\theta}_{app,L}, \hat{\theta}_{app,U})$; for the 95% confidence interval, they are typically calculated by the 2.5th and 97.5th percentiles of the bootstrap distribution.

3. Calculate the bias estimate by optimism, $\hat{\delta} = \hat{\theta}_{app} - \hat{\theta}$.

4. Then, the location-shifted bootstrap confidence interval is computed as $(\hat{\theta}_{app,L} - \hat{\delta}, \hat{\theta}_{app,U} - \hat{\delta})$.

Note that the advantage of this method is the simplicity and low computational burden of calculating the confidence limits. It only requires the bootstrap confidence interval by $\hat{\theta}_{app}$ and the bias estimate $\hat{\delta}$. These quantities can be obtained within the bootstrap processes of the optimism correction methods, and additional burdensome computations are not needed. Adjusting the location of the apparent bootstrap confidence interval is straightforward, but the location-shifted confidence interval is justified by the large-sample theory, because the apparent bootstrap confidence interval becomes a valid confidence interval asymptotically [25,26], and the bias also converges to 0 [2].

However, the apparent bootstrap confidence interval only quantifies the variability of the apparent predictive measure $\hat{\theta}_{app}$. The optimism-corrected measure $\hat{\theta}$ generally has a larger variability related to the variability of the optimism correction quantity $\hat{\delta}$ and the correlation between $\hat{\theta}_{app}$ and $\hat{\delta}$. Therefore, it can underestimate the statistical variability and have undercoverage properties under moderate sample settings.

*3.2 Two-stage bootstrap confidence interval*

To address the undercoverage properties, the variability of the optimism measures $\hat{\delta}$ and the correlations between $\hat{\theta}_{app}$ and $\hat{\delta}$ should be adequately considered. However, the correlations between their constituent components are quite complicated and are difficult to assess adequately. Thus, numerical approaches are also effective for assessing their variabilities simultaneously. We propose the following two-stage bootstrap approach that aims to directly obtain bootstrap distributions of the optimism-corrected statistics.

*Algorithm 2 (Two-stage bootstrap confidence interval)*

1. Generate $B$ bootstrap samples by resampling with replacement from the original population.

2. Develop a multivariable prediction model for each bootstrap sample, and calculate the optimism-corrected predictive measures for the $B$ bootstrap samples $\hat{\theta}_1, \hat{\theta}_2, \ldots, \hat{\theta}_B$, using the Harrell's bias correction, 0.632, or 0.632+ method, by performing bootstrap resampling from the bootstrap sample. We can obtain the bootstrap distributions of the optimism-corrected predictive measures by $\hat{\theta}_1, \hat{\theta}_2, \ldots, \hat{\theta}_B$.

3. Then, compute the bootstrap confidence interval from the bootstrap samples $\hat{\theta}_1, \hat{\theta}_2, \ldots, \hat{\theta}_B$; for the 95% confidence interval, they are typically calculated by the 2.5th and 97.5th percentiles of the bootstrap distribution.

The two-stage bootstrap confidence interval adequately addresses the variabilities of the optimism-corrected measures themselves that involve the correlations mentioned above. In the numerical evaluations in Section 4, the two-stage bootstrap methods provided wider confidence intervals in general and showed better coverage properties compared with the apparent bootstrap confidence interval. In addition, the two-stage bootstrap directly corresponds to the formal bootstrap confidence intervals of the Harrell's bias correction, 0.632, and 0.632+ methods, thus their asymptotic validities are assured theoretically [25,26].

However, a difficulty of this approach is the computational burden. Many studies recommend performing more than 1000 resamplings for calculating bootstrap confidence intervals, so if we conduct 2000 bootstrap resamplings for both of the two-stage bootstraps, $2000 \times 2000 = 4000000$ iterative computations are needed to construct the multivariable prediction models, requiring enormous computational time. However, the performance of computers is continually increasing, and therefore these tasks might not be difficult in the near future. At present, an appropriate solution is to collaborate with statisticians who have access to high-performance computer systems. If requested, our group is willing to engage in collaborations on valuable research projects.

The R package `predboot`, which can conduct the two proposed methods for logistic regression models with the ML, ridge, and lasso estimations, is available at https://github.com/nomahi/predboot. Also, examples of `predboot` codes for implementing the estimations are available in e-Appendix A.

## 4. Applications

To illustrate the effectiveness of the proposed methods in practice, we present their applications to a real-world dataset, namely that of the GUSTO-I trial [13,14]. The GUSTO-I dataset has been adopted by many performance evaluation studies of multivariable prediction models [7,28-30], and here we specifically used the West region dataset. GUSTO-I was a comparative clinical trial that assessed four treatment strategies for acute myocardial infarction. Here we adopted death within 30 days as the outcome variable. There were 17 covariates: two variables (height and weight) are continuous, one (smoking) is ordinal, and the remaining 14 (age, gender, diabetes, hypotension, tachycardia, high risk, shock, no relief of chest pain, previous myocardial infraction, hypertension history, hypercholesterolemia, previous angina pectoris, family history of myocardial infarction, and ST elevation in >4 leads) are binary; age was dichotomized at 65 years old. For smoking, which is a three-category variable (current smokers, ex-smokers, and never-smokers), we generated two dummy variables (ex-smokers vs. never-smokers, and current smokers vs. never-smokers) and used them in the analyses. The clinical trial dataset can be downloaded from http://www.clinicalprediction models.org.

We considered two modelling strategies: (1) 8-predictor models (age > 65 years, female gender, diabetes, hypotension, tachycardia, high risk, shock, and no relief of chest pain), which were adopted in several previous studies [7,28], and (2) 17-predictor models, which included all the variables mentioned above. The EPVs for these models were 16.9 and 7.5, respectively. For the two modelling strategies, we constructed multivariable logistic prediction models by ML estimation and the two shrinkage penalized regression approaches, the ridge and lasso regressions. We applied the proposed methods to these prediction models, and calculated the $C$-statistics and the bootstrap-based confidence intervals. The number of bootstrap resamplings was consistently set to 2000; for the two-stage bootstrap methods, the total resampling number was $2000 \times 2000 = 4000000$. We assessed the Monte Carlo errors of the two-stage bootstrap confidence intervals by varying the resampling numbers, and the results are presented in e-Appendix B.

For the 8-predictor models, the results are presented in Table 2. The corrected optimisms from the apparent $C$-statistics were 0.009 for ML estimation and 0.007-0.008 for the ridge and lasso regressions. The 95% confidence intervals by DeLong's method and apparent bootstrap were located around the optimism-uncorrected $C$-statistic, and was a little influenced by the biases. For the ML estimation, the two proposed methods provided optimism-corrected confidence intervals, as expected. Further, for the ridge estimation, the locations of the 95% confidence intervals moved upward, i.e., the bootstrap distributions of the optimised-corrected *C*-statistics moved upward. For the lasso regression, the lower confidence limits moved upward compared with the location-shifted bootstrap confidence intervals. On the other hand, the upper confidence limits moved downward, i.e., the bootstrap distributions were limited to narrower ranges. The results were not very different between the three optimism correction methods, i.e., the Harrell's bias correction, 0.632, and 0.632+ methods.

For the 17-predictor models, the results are presented in Table 3. The corrected optimisms from the apparent *C*-statistics were generally larger than those from the 8-predictor models, and became 0.021-0.022 for ML estimation and 0.018-0.019 for ridge and lasso regressions. The DeLong's and apparent bootstrap confidence intervals could be influenced by biases. The location-shifted bootstrap confidence intervals would correct the relatively large biases. For the two-stage bootstrap confidence intervals, the overall results were similar to those obtained with the ML estimation; they provided slightly wider confidence intervals compared with the location-shifted bootstrap confidence intervals. For the ridge estimation, the locations of the lower confidence limits moved upward, as did the bootstrap distributions of the optimism-corrected *C*-statistics. These results indicate that the AUCs became larger. For the lasso regression, the lower confidence limits were very different from those of the location-shifted bootstrap confidence intervals, but the upper confidence limits moved downward. The confidence intervals indicate that the standard errors of the *C*-statistics became smaller but the prediction performances became worse as a result of the strong shrinkage with the lasso regressions. For the 17-predictor models, the results were also not very different among the three optimism correction methods, namely the Harrell, 0.632, and 0.632+ methods.

# 5. Simulation experiments

## *5.1 Simulation settings*

To evaluate the performances of the proposed methods, we conducted simulation studies based on the GUSTO-I trial [13,14]. We explain the settings of simulation studies following the ADEMP structure of Morris et al. [31] as follows.

*Aims*: The aims of the simulation studies were to assess the validities of the proposed methods and to compare their performances with those of the conventional standard methods.

*Data-generating mechanisms*: We considered a certain range of conditions with various factors that can affect predictive performance: the events per variable (EPV) (1, 3, 5, 7, 10, 20, and 40), the expected event fraction (0.125 and 0.0625), the number of candidate predictors (8 and 17 variables), and the regression coefficients of the predictor variables (two scenarios, as explained below). A total of 56 scenarios covering all combinations of these settings were investigated. The settings of the EPV and the event fraction were based on those used in previous studies [7,32]. For the regression coefficients of the predictor variables (except for the intercept $\beta_0$), we considered two scenarios: one fixed to the ML estimate for the GUSTO-I dataset (coefficient type 1) and the other fixed to the lasso estimate for the same dataset (coefficient type 2). With coefficient type 1, all the predictors had some effect on the risk of events, while with coefficient type 2, some of the predictor effects were null and the others were relatively small compared with scenario 1; the two scenarios are considered as plausible settings estimated from the real world dataset using valid estimation methods. The intercept $\beta_0$ was set to properly adjust the event fractions. We also conducted additional simulations, shown in in e-Appendix C, that varied the expected event fractions to 0.25 and 0.50. Also, the sample size of the derivation cohort $n$ was determined as follows: (the number of candidate predictor variables × EPV) / (expected event fraction). The predictor variables were generated as random numbers based on the parameters estimated from the GUSTO-I dataset; for the details of the covariate information, see Section 4. Three continuous variables (height, weight, and age) were generated from a multivariate normal distribution with the same mean vector and covariance matrix as in the GUSTO-I data. For smoking, an ordinal variable, random numbers were generated from a multinomial distribution using the same proportions as in the GUSTO-I data; this variable was converted to two dummy variables before being incorporated into the prediction models. In addition, the remaining binary variables were generated from a multivariate binomial distribution [33] using the same marginal probabilities and correlation coefficients estimated from the GUSTO-I dataset. We used the `mipfp` package [34] to generate the correlated binomial variables. We also conducted additional simulations, shown in e-Appendix C, that used resamples of the covariates from the original GUSTO-I dataset and that completely preserved the correlation structure. The event occurrence probability $\pi_i$ ($i = 1, 2, \ldots, n$) was determined based on the generated predictor variables $x_i$ and the logistic regression model $\pi_i = 1/(1 + \exp(-\beta^T x_i))$. The outcome variable $y_i$ was generated from a Bernoulli distribution with a success probability $\pi_i$.

*Estimands*: We evaluated the prediction performance of the *C*-statistic, which is the most popular discriminant measure for clinical prediction models. The estimand was set to the empirical AUC of the ROC curve for 500000 independently generated external test observations. We estimated the AUC using the apparent *C*-statistic and the bootstrap-based optimism-corrected *C*-statistic.

*Methods*: We evaluated the coverage performances and precisions of 95% confidence intervals for the AUC by (1) DeLong's confidence interval [8], (2) the apparent bootstrap confidence interval, (3-5) the location-shifted bootstrap confidence intervals by the Harrell, 0.632, and 0.632+ methods, and (6-8) the two-stage bootstrap confidence intervals by the Harrell, 0.632, and 0.632+ methods. The number of bootstrap resamplings was consistently set to 1000; for the two-stage bootstrap methods, the total resampling number was $1000 \times 1000 = 1000000$. The multivariable prediction model was constructed by ML estimation.

*Performance measures*: The coverage rates of the AUC and the expected widths were adopted as evaluation measures. We conducted 1000 simulations under each scenario, and empirical measures of these quantities were evaluated.

## 5.2 Results

The results of the simulations are presented in Figures 1-3 using the nested-loop plots of Rücker and Schwarzer [35]. Specifically, the results of the coverage rates of 95% confidence intervals are shown in Figure 1, and the expected widths of these confidence intervals in Figure 2. First, under most of the 56 scenarios, the apparent bootstrap confidence interval showed marked undercoverage properties. The coverage rates were especially small when EPV was small and/or the number of predictor variables was large. The DeLong's confidence interval had similar trends and showed undercoverage properties, but the actual coverage rate was relatively large compared with the apparent bootstrap confidence interval. These results indicate that these naive methods lacking optimism corrections usually misestimate the statistical variability of prediction accuracy measures.

The proposed methods showed clearly more favourable coverage performances. For the location-shifted bootstrap confidence intervals, all three methods performed well and their coverage rates were around the nominal level (95%) when EPV was relatively large ($\geq 10$). However, they showed minor undercoverage properties in general, and the trend was fairly strong under relatively small EPV settings. The undercoverage properties were pronounced under especially small EPV settings (EPV = 1, 3). The undercoverage properties would be caused by the fact that the location-shifted bootstrap confidence intervals only quantify the statistical errors by the apparent bootstrap confidence interval, and may underestimate the total statistical variabilities as mentioned in Section 3. However, these results show that the estimated variation with the apparent bootstrap confidence interval certainly quantifies the actual statistical variabilities under relatively large sample sizes, although they can be easily computed by the bootstrap outputs using the ordinary optimism correction methods.

In addition, for the two-stage bootstrap confidence intervals, the coverage rates were superior to the location-shifted bootstrap confidence intervals in general. Under all the scenarios except for the settings with EPV = 1, the realized coverage rates were around the nominal level (95%), and their expected widths were slightly larger than the apparent bootstrap confidence intervals (i.e., equal to those of the location-shifted confidence intervals). The differences of coverage rates with the location-shifted confidence intervals were markedly under those of the EPV $\leq$ 10 settings. However, under especially small EPV settings (EPV = 1), the coverage performances were not very good, and the confidence interval based on Harrell's method showed undercoverage properties. That based on the 0.632 estimator also demonstrated undercoverage under a few scenarios. In contrast, the confidence interval using the 0.632+ estimator showed valid coverage rates; this estimator is therefore recommended under small EPV settings. Comparing coefficient types 1 and 2, the general performances of the three methods were comparable, but the coverage probabilities were generally larger under the former settings. In summary, the two-stage bootstrap methods provided values that were nearly identical to the nominal level under most of these settings. This indicates that the two-stage bootstrap can adequately assess the statistical variabilities of the optimism-corrected predictive measures.

## 6. Discussion

In developing multivariable prediction models, bootstrapping methods for internal validations of discriminant and calibration measures have been increasingly used in recent clinical studies [15,36]. Although most published papers have presented the optimism-uncorrected confidence intervals (e.g., DeLong's method for the *C*-statistic [8]), they sometimes provide inaccurate and misleading evidence. In this paper, our simulations clearly showed the inadequacy of the naïve methods that do not consider optimism, and their use is not recommended in practice. Adequate alternative methods should be used instead.

In this article, we proposed two effective methods to construct the confidence intervals to address this important issue. The most highly recommended approach is to use the two-stage bootstrap methods. These adequately reflect the statistical variability of the optimism-corrected prediction measures, and they provide appropriate confidence intervals as shown in the simulation studies. However, one difficulty is that they have a heavy computational burden. Their application in practice will be difficult if analysts cannot access a high-performance computer that can conduct parallel computations. However, the prices of these high-performance machines have been gradually becoming more reasonable, and they will not be considered to be special tools in the future. A current solution is to collaborate with statisticians who have access to high-performance computer systems. We are willing to collaborate on valuable research projects upon request. Also, the computations can be easily implemented using the R package `predboot` (https://github.com/nomahi/predboot).

An alternative approach is the use of location-shifted bootstrap confidence intervals. Although these may underestimate statistical variabilities, as mentioned in Section 3.1, the actual coverage rates of true predictive accuracy measures approach the nominal level when the sample sizes are reasonably large. In our simulation studies, the coverage performances were favourable under EPV ≥ 20, and would be acceptable even under EPV = 10. They were certainly better than the Delong's and apparent bootstrap confidence intervals. In addition, the location-shifted bootstrap confidence intervals can be computed by the outputs of bootstrap algorithms for calculating the optimism-corrected prediction measures, and a pragmatic advantage is that no additional computational burdens are needed.

In future methodological studies, alternative effective computational methods might be developed, such as those of LeDell et al. [9], which combine analytical and Monte Carlo approaches. These are relevant issues for further research. Also, the proposed methods can be applied to other optimism correction methods by bootstrap, e.g., the Efron-Gong [37] method. Further theoretical discussions comparing the optimism correction methods, especially the Efron-Gong [37] and 0.632 methods, would be future relevant issues. In summary, for the evaluations of predictive accuracies of multivariable prediction models, the naïve confidence intervals are no longer recommended, and appropriate methods should be adopted in practice. The proposed methods in this article certainly provide a pragmatic solution and can serve as effective practical tools.

**Acknowledgements**

This study was supported by a Grant-in-Aid for Scientific Research from the Japan Society for the Promotion of Science (Grant number: JP19H04074).

**Table 1**. *C*-statistics and 95% confidence intervals (C.I.) for the GUSTO-I trial dataset for the 8-variable models.

|  | ML estimation | Ridge estimation | Lasso estimation |
|---|---|---|---|
| Apparent *C*-statistic | | | |
|   DeLong's C.I. | 0.819 (0.783, 0.854) | 0.819 (0.784, 0.855) | 0.819 (0.787, 0.857) |
|   Apparent bootstrap C.I. | 0.819 (0.788, 0.858) | 0.819 (0.787, 0.858) | 0.819 (0.787, 0.857) |
| Harrell's bias correction | | | |
|   Location-shifted bootstrap C.I. | 0.810 (0.779, 0.849) | 0.811 (0.779, 0.850) | 0.811 (0.779, 0.849) |
|   Two-stage bootstrap C.I. | 0.810 (0.777, 0.850) | 0.811 (0.787, 0.857) | 0.811 (0.784, 0.839) |
| 0.632 estimator | | | |
|   Location-shifted bootstrap C.I. | 0.810 (0.779, 0.849) | 0.812 (0.780, 0.851) | 0.811 (0.779, 0.849) |
|   Two-stage bootstrap C.I. | 0.810 (0.777, 0.850) | 0.812 (0.788, 0.857) | 0.811 (0.784, 0.840) |
| 0.632+ estimator | | | |
|   Location-shifted bootstrap C.I. | 0.810 (0.779, 0.849) | 0.812 (0.780, 0.851) | 0.811 (0.779, 0.849) |
|   Two-stage bootstrap C.I. | 0.810 (0.777, 0.850) | 0.812 (0.788, 0.857) | 0.811 (0.784, 0.840) |

**Table 2**. *C*-statistics and 95% confidence intervals (C.I.) for the GUSTO-I trial dataset for the 17-variable models.

|  | ML estimation | Ridge estimation | Lasso estimation |
|---|---|---|---|
| Apparent *C*-statistic |  |  |  |
|   DeLong's C.I. | 0.832 (0.796, 0.867) | 0.831 (0.795, 0.866) | 0.831 (0.795, 0.866) |
|   Apparent bootstrap C.I. | 0.832 (0.803, 0.874) | 0.831 (0.804, 0.873) | 0.831 (0.804, 0.873) |
| Harrell's bias correction |  |  |  |
|   Location-shifted bootstrap C.I. | 0.811 (0.782, 0.853) | 0.812 (0.785, 0.854) | 0.813 (0.786, 0.855) |
|   Two-stage bootstrap C.I. | 0.811 (0.782, 0.858) | 0.812 (0.794, 0.856) | 0.813 (0.786, 0.848) |
| 0.632 estimator |  |  |  |
|   Location-shifted bootstrap C.I. | 0.811 (0.782, 0.853) | 0.813 (0.786, 0.855) | 0.813 (0.786, 0.855) |
|   Two-stage bootstrap C.I. | 0.811 (0.782, 0.857) | 0.813 (0.794, 0.856) | 0.813 (0.785, 0.848) |
| 0.632+ estimator |  |  |  |
|   Location-shifted bootstrap C.I. | 0.810 (0.781, 0.852) | 0.812 (0.785, 0.854) | 0.813 (0.786, 0.855) |
|   Two-stage bootstrap C.I. | 0.810 (0.781, 0.856) | 0.812 (0.793, 0.856) | 0.813 (0.785, 0.848) |

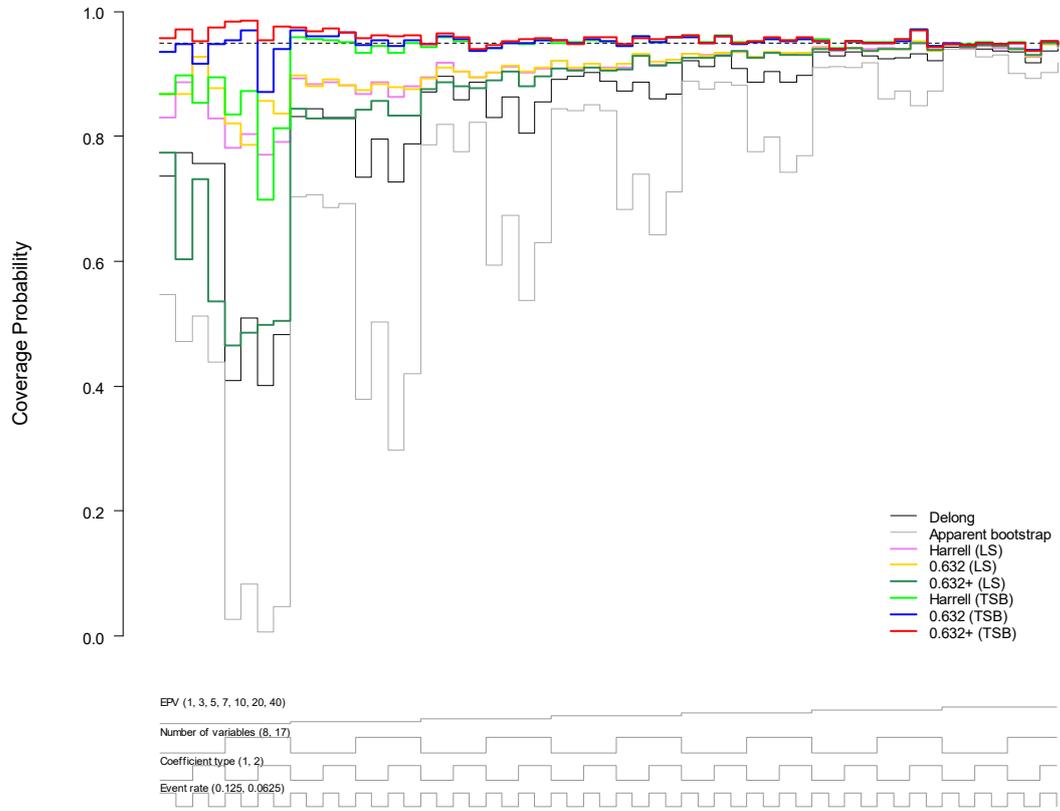

**Figure 1.** The coverage rates of 95% confidence intervals in the simulation studies by the DeLong method, the apparent bootstrap confidence interval, the location-shifted (LS) confidence intervals and the two-stage bootstrap (TSB) confidence intervals for the Harrell, 0.632, and 0.632+ methods.

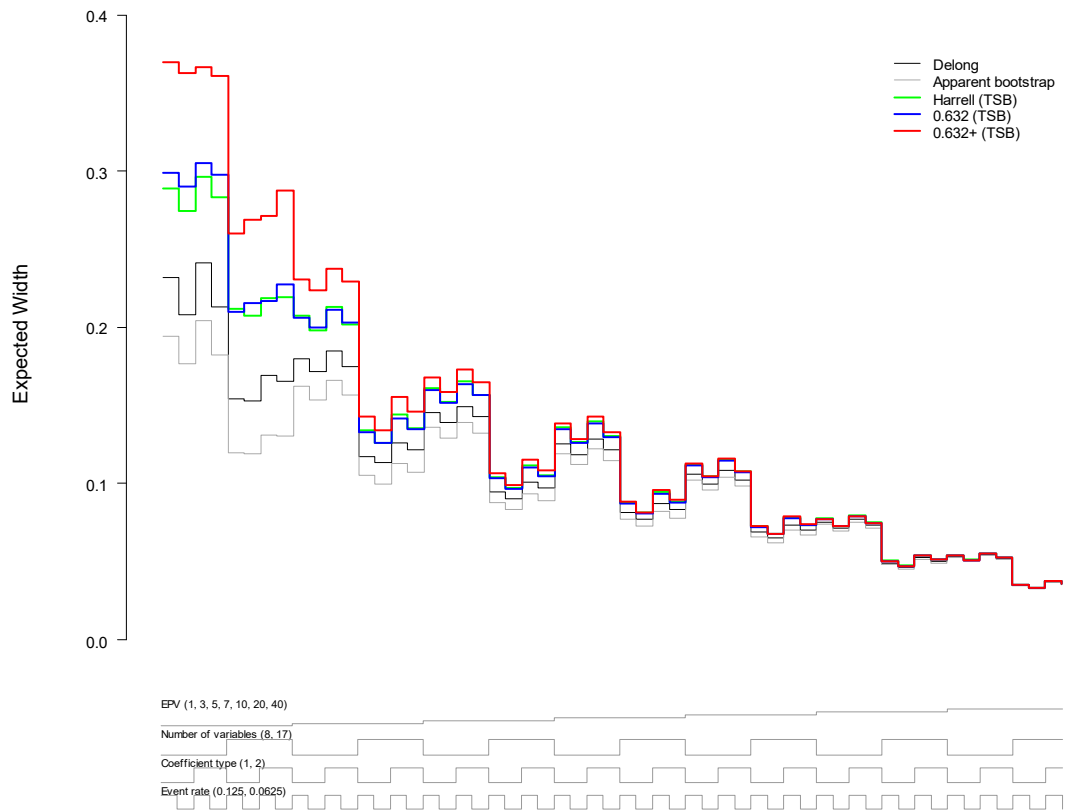

**Figure 2.** The expected widths of 95% confidence intervals in the simulation studies by the DeLong method, the apparent bootstrap confidence interval, and the two-stage bootstrap (TSB) confidence intervals for the Harrell, 0.632, and 0.632+ methods; the expected widths of the location-shifted confidence intervals are consistent with those of the apparent bootstrap confidence interval.

# Confidence intervals of prediction accuracy measures for multivariable prediction models based on the bootstrap-based optimism correction methods

Hisashi Noma[1], Tomohiro Shinozaki[2], Katsuhiro Iba[3,4], Satoshi Teramukai[5] and Toshi A. Furukawa[6]

[1] Department of Data Science, The Institute of Statistical Mathematics, Tokyo, Japan
[2] Department of Information and Computer Technology, Faculty of Engineering, Tokyo University of Science, Tokyo, Japan
[3] Department of Statistical Science, School of Multidisciplinary Sciences, The Graduate University for Advanced Studies, Tokyo, Japan
[4] Office of Biostatistics, Department of Biometrics, Headquarters of Clinical Development, Otsuka Pharmaceutical Co., Ltd., Tokyo, Japan
[5] Department of Biostatistics, Graduate School of Medical Science, Kyoto Prefectural University of Medicine, Kyoto, Japan
[6] Departments of Health Promotion and Human Behavior and of Clinical Epidemiology, Kyoto University Graduate School of Medicine/School of Public Health, Kyoto, Japan

## e-Appendix A: R example codes

```
# Installation of the predboot package
require(devtools)
devtools::install_github("nomahi/predboot")

# Load the example dataset
library(predboot)
?predboot          # Help file
data(exdata)       # A hypothetical simulated cohort dataset

# Location-shifted bootstrap CI for ML estimation
pred.ML(Y ~ A65 + SEX + DIA + HYP + HRT + HIG + SHO + TTR,
data=exdata, B=1000)    # 8 variables model
pred.ML(Y ~ A65 + SEX + DIA + HYP + HRT + HIG + SHO + TTR +
PMI + HEI + WEI + HTN + SMK1 + SMK2 + LIP + PAN + FAM + ST4,
data=exdata, B=1000)    # 17 variables model
```

```r
# Two-stage bootstrap CI for ML estimation
pred.ML2(Y ~ A65 + SEX + DIA + HYP + HRT + HIG + SHO + TTR,
data=exdata, B=1000)     # 8 variables model
pred.ML2(Y ~ A65 + SEX + DIA + HYP + HRT + HIG + SHO + TTR
+ PMI + HEI + WEI + HTN + SMK1 + SMK2 + LIP + PAN + FAM +
ST4, data=exdata, B=1000)     # 17 variables model

# Location-shifted bootstrap CI for ridge estimation
pred.ridge(Y ~ A65 + SEX + DIA + HYP + HRT + HIG + SHO + TTR,
data=exdata, B=1000)     # 8 variables model
pred.ridge(Y ~ A65 + SEX + DIA + HYP + HRT + HIG + SHO + TTR
+ PMI + HEI + WEI + HTN + SMK1 + SMK2 + LIP + PAN + FAM +
ST4, data=exdata, B=1000)     # 17 variables model

# Two-stage bootstrap CI for ridge estimation
pred.ridge2(Y ~ A65 + SEX + DIA + HYP + HRT + HIG + SHO +
TTR, data=exdata, B=1000)     # 8 variables model
pred.ridge2(Y ~ A65 + SEX + DIA + HYP + HRT + HIG + SHO +
TTR + PMI + HEI + WEI + HTN + SMK1 + SMK2 + LIP + PAN + FAM
+ ST4, data=exdata, B=1000)     # 17 variables model

# Location-shifted bootstrap CI for lasso estimation
pred.lasso(Y ~ A65 + SEX + DIA + HYP + HRT + HIG + SHO + TTR,
data=exdata, B=1000)     # 8 variables model
pred.lasso(Y ~ A65 + SEX + DIA + HYP + HRT + HIG + SHO + TTR
+ PMI + HEI + WEI + HTN + SMK1 + SMK2 + LIP + PAN + FAM +
ST4, data=exdata, B=1000)     # 17 variables model

# Two-stage bootstrap CI for lasso estimation
pred.lasso2(Y ~ A65 + SEX + DIA + HYP + HRT + HIG + SHO +
TTR, data=exdata, B=1000)     # 8 variables model
```

```
pred.lasso2(Y ~ A65 + SEX + DIA + HYP + HRT + HIG + SHO +
TTR + PMI + HEI + WEI + HTN + SMK1 + SMK2 + LIP + PAN + FAM
+ ST4, data=exdata, B=1000)   # 17 variables model
```

For more detail information, please see the help files of the `predboot` package and the web page (https://github.com/nomahi/predboot).

### e-Appendix B: Assessments of Monte Carlo errors of two-stage bootstrap confidence intervals

To assess the Monte Carlo errors of the two-stage bootstrap confidence intervals, we created 5000 different two-stage bootstrap confidence intervals for the GUSTO-I dataset in Section 4 by varying the numbers of inner and outer bootstraps ($B_{inner}$ and $B_{outer}$, respectively). The estimation method was ML and the 8-predictor model was adopted. We present the summary statistics (mean, SD, and quantiles) of 5000 confidence limits derived using the Harrell, 0.632, and 0.632+ methods in e-Table 1-6. The number of inner bootstraps was not strongly influenced by the Monte Carlo errors, and the resultant confidence limits changed only minimally, even with inner bootstrap numbers of 100 or 200. This is because the inner bootstrap was used for point estimation of the optimism bias. In contrast with the above results, the number of outer bootstraps was directly influenced by the Monte Carlo errors. The outer bootstraps were used for quantile estimations of the bootstrap distributions, so a sufficient number of replications is required, ideally more than 1000 in practice. To strictly control Monte Carlo errors, we adopted 1000 or 2000 as the numbers of bootstrap resampling in the numerical analyses in Sections 4 and 5, but the number of inner bootstraps could be reduced for computational efficiency.

e-Table 1. Distribution of the lower 95% confidence limit of the AUC for the GUSTO-I dataset using the 8-predictor model with the two-stage bootstrap confidence interval through 5000 replications: the Harrell's bias correction.

| $B_{outer}$ | $B_{inner}$ | Mean | SD | Quantiles | | | | |
|---|---|---|---|---|---|---|---|---|
| | | | | 2.5% | 25% | 50% | 75% | 97.5% |
| 500 | 100 | 0.778 | 0.002 | 0.773 | 0.776 | 0.778 | 0.779 | 0.782 |
| | 200 | 0.778 | 0.002 | 0.773 | 0.776 | 0.778 | 0.779 | 0.782 |
| | 300 | 0.778 | 0.002 | 0.773 | 0.776 | 0.778 | 0.779 | 0.782 |
| | 400 | 0.778 | 0.002 | 0.773 | 0.776 | 0.778 | 0.779 | 0.782 |
| | 500 | 0.778 | 0.002 | 0.773 | 0.776 | 0.778 | 0.779 | 0.782 |
| | 1000 | 0.778 | 0.002 | 0.773 | 0.776 | 0.778 | 0.779 | 0.782 |
| | 1500 | 0.778 | 0.002 | 0.773 | 0.776 | 0.778 | 0.779 | 0.782 |
| | 2000 | 0.778 | 0.002 | 0.773 | 0.776 | 0.778 | 0.779 | 0.782 |
| 1000 | 100 | 0.777 | 0.002 | 0.774 | 0.776 | 0.777 | 0.779 | 0.781 |
| | 200 | 0.778 | 0.002 | 0.774 | 0.776 | 0.778 | 0.779 | 0.781 |
| | 300 | 0.778 | 0.002 | 0.774 | 0.776 | 0.778 | 0.779 | 0.781 |
| | 400 | 0.778 | 0.002 | 0.774 | 0.776 | 0.778 | 0.779 | 0.781 |
| | 500 | 0.778 | 0.002 | 0.774 | 0.776 | 0.778 | 0.779 | 0.781 |
| | 1000 | 0.778 | 0.002 | 0.774 | 0.776 | 0.778 | 0.779 | 0.781 |
| | 1500 | 0.778 | 0.002 | 0.774 | 0.776 | 0.778 | 0.779 | 0.781 |
| | 2000 | 0.778 | 0.002 | 0.774 | 0.776 | 0.778 | 0.779 | 0.781 |
| 1500 | 100 | 0.777 | 0.001 | 0.775 | 0.776 | 0.777 | 0.778 | 0.780 |
| | 200 | 0.777 | 0.001 | 0.775 | 0.777 | 0.778 | 0.778 | 0.780 |
| | 300 | 0.778 | 0.001 | 0.775 | 0.777 | 0.778 | 0.778 | 0.780 |
| | 400 | 0.778 | 0.001 | 0.775 | 0.777 | 0.778 | 0.778 | 0.780 |
| | 500 | 0.778 | 0.001 | 0.775 | 0.777 | 0.778 | 0.778 | 0.780 |
| | 1000 | 0.778 | 0.001 | 0.775 | 0.777 | 0.778 | 0.779 | 0.780 |
| | 1500 | 0.778 | 0.001 | 0.775 | 0.777 | 0.778 | 0.778 | 0.780 |
| | 2000 | 0.778 | 0.001 | 0.775 | 0.777 | 0.778 | 0.779 | 0.780 |
| 2000 | 100 | 0.777 | 0.001 | 0.775 | 0.777 | 0.777 | 0.778 | 0.780 |
| | 200 | 0.777 | 0.001 | 0.775 | 0.777 | 0.777 | 0.778 | 0.780 |
| | 300 | 0.777 | 0.001 | 0.775 | 0.777 | 0.777 | 0.778 | 0.780 |
| | 400 | 0.777 | 0.001 | 0.775 | 0.777 | 0.777 | 0.778 | 0.780 |
| | 500 | 0.778 | 0.001 | 0.775 | 0.777 | 0.778 | 0.778 | 0.780 |
| | 1000 | 0.778 | 0.001 | 0.775 | 0.777 | 0.778 | 0.778 | 0.780 |
| | 1500 | 0.778 | 0.001 | 0.775 | 0.777 | 0.778 | 0.778 | 0.780 |
| | 2000 | 0.778 | 0.001 | 0.775 | 0.777 | 0.778 | 0.778 | 0.780 |

e-Table 2. Distribution of the upper 95% confidence limit of the AUC for the GUSTO-I dataset using the 8-predictor model with the two-stage bootstrap confidence interval through 5000 replications: the Harrell's bias correction.

| $B_{outer}$ | $B_{inner}$ | Mean | SD | Quantiles | | | | |
|---|---|---|---|---|---|---|---|---|
| | | | | 2.5% | 25% | 50% | 75% | 97.5% |
| 500 | 100 | 0.850 | 0.002 | 0.846 | 0.848 | 0.850 | 0.851 | 0.854 |
| | 200 | 0.850 | 0.002 | 0.846 | 0.848 | 0.850 | 0.851 | 0.854 |
| | 300 | 0.850 | 0.002 | 0.846 | 0.848 | 0.850 | 0.851 | 0.854 |
| | 400 | 0.850 | 0.002 | 0.846 | 0.848 | 0.850 | 0.851 | 0.854 |
| | 500 | 0.850 | 0.002 | 0.846 | 0.848 | 0.850 | 0.851 | 0.854 |
| | 1000 | 0.850 | 0.002 | 0.846 | 0.848 | 0.849 | 0.851 | 0.854 |
| | 1500 | 0.850 | 0.002 | 0.846 | 0.848 | 0.849 | 0.851 | 0.854 |
| | 2000 | 0.850 | 0.002 | 0.846 | 0.848 | 0.849 | 0.851 | 0.854 |
| 1000 | 100 | 0.850 | 0.001 | 0.847 | 0.849 | 0.850 | 0.851 | 0.853 |
| | 200 | 0.850 | 0.001 | 0.847 | 0.849 | 0.850 | 0.851 | 0.853 |
| | 300 | 0.850 | 0.001 | 0.847 | 0.849 | 0.850 | 0.851 | 0.852 |
| | 400 | 0.850 | 0.001 | 0.847 | 0.849 | 0.850 | 0.851 | 0.852 |
| | 500 | 0.850 | 0.001 | 0.847 | 0.849 | 0.850 | 0.851 | 0.853 |
| | 1000 | 0.850 | 0.001 | 0.847 | 0.849 | 0.850 | 0.851 | 0.853 |
| | 1500 | 0.850 | 0.001 | 0.847 | 0.849 | 0.850 | 0.851 | 0.853 |
| | 2000 | 0.850 | 0.001 | 0.847 | 0.849 | 0.850 | 0.851 | 0.853 |
| 1500 | 100 | 0.850 | 0.001 | 0.848 | 0.849 | 0.850 | 0.851 | 0.852 |
| | 200 | 0.850 | 0.001 | 0.848 | 0.849 | 0.850 | 0.851 | 0.852 |
| | 300 | 0.850 | 0.001 | 0.848 | 0.849 | 0.850 | 0.851 | 0.852 |
| | 400 | 0.850 | 0.001 | 0.848 | 0.849 | 0.850 | 0.851 | 0.852 |
| | 500 | 0.850 | 0.001 | 0.848 | 0.849 | 0.850 | 0.851 | 0.852 |
| | 1000 | 0.850 | 0.001 | 0.848 | 0.849 | 0.850 | 0.851 | 0.852 |
| | 1500 | 0.850 | 0.001 | 0.848 | 0.849 | 0.850 | 0.851 | 0.852 |
| | 2000 | 0.850 | 0.001 | 0.848 | 0.849 | 0.850 | 0.851 | 0.852 |
| 2000 | 100 | 0.850 | 0.001 | 0.848 | 0.849 | 0.850 | 0.851 | 0.852 |
| | 200 | 0.850 | 0.001 | 0.848 | 0.849 | 0.850 | 0.851 | 0.852 |
| | 300 | 0.850 | 0.001 | 0.848 | 0.849 | 0.850 | 0.850 | 0.852 |
| | 400 | 0.850 | 0.001 | 0.848 | 0.849 | 0.850 | 0.850 | 0.852 |
| | 500 | 0.850 | 0.001 | 0.848 | 0.849 | 0.850 | 0.850 | 0.852 |
| | 1000 | 0.850 | 0.001 | 0.848 | 0.849 | 0.850 | 0.850 | 0.852 |
| | 1500 | 0.850 | 0.001 | 0.848 | 0.849 | 0.850 | 0.850 | 0.852 |
| | 2000 | 0.850 | 0.001 | 0.848 | 0.849 | 0.850 | 0.850 | 0.852 |

**e-Table 3.** Distribution of the lower 95% confidence limit of the AUC for the GUSTO-I dataset using the 8-predictor model with the two-stage bootstrap confidence interval through 5000 replications: the 0.632 method.

| $B_{outer}$ | $B_{inner}$ | Mean | SD | Quantiles | | | | |
|---|---|---|---|---|---|---|---|---|
| | | | | 2.5% | 25% | 50% | 75% | 97.5% |
| 500 | 100 | 0.778 | 0.002 | 0.773 | 0.776 | 0.778 | 0.780 | 0.782 |
| | 200 | 0.778 | 0.002 | 0.773 | 0.777 | 0.778 | 0.780 | 0.782 |
| | 300 | 0.778 | 0.002 | 0.773 | 0.777 | 0.778 | 0.780 | 0.782 |
| | 400 | 0.778 | 0.002 | 0.773 | 0.777 | 0.778 | 0.780 | 0.782 |
| | 500 | 0.778 | 0.002 | 0.773 | 0.777 | 0.778 | 0.780 | 0.783 |
| | 1000 | 0.778 | 0.002 | 0.773 | 0.777 | 0.778 | 0.780 | 0.783 |
| | 1500 | 0.778 | 0.002 | 0.773 | 0.777 | 0.778 | 0.780 | 0.783 |
| | 2000 | 0.778 | 0.002 | 0.773 | 0.777 | 0.778 | 0.780 | 0.783 |
| 1000 | 100 | 0.778 | 0.002 | 0.774 | 0.777 | 0.778 | 0.779 | 0.781 |
| | 200 | 0.778 | 0.002 | 0.774 | 0.777 | 0.778 | 0.779 | 0.781 |
| | 300 | 0.778 | 0.002 | 0.774 | 0.777 | 0.778 | 0.779 | 0.781 |
| | 400 | 0.778 | 0.002 | 0.774 | 0.777 | 0.778 | 0.779 | 0.781 |
| | 500 | 0.778 | 0.002 | 0.775 | 0.777 | 0.778 | 0.779 | 0.781 |
| | 1000 | 0.778 | 0.002 | 0.775 | 0.777 | 0.778 | 0.779 | 0.781 |
| | 1500 | 0.778 | 0.002 | 0.775 | 0.777 | 0.778 | 0.779 | 0.781 |
| | 2000 | 0.778 | 0.002 | 0.775 | 0.777 | 0.778 | 0.779 | 0.781 |
| 1500 | 100 | 0.778 | 0.001 | 0.775 | 0.777 | 0.778 | 0.779 | 0.780 |
| | 200 | 0.778 | 0.001 | 0.775 | 0.777 | 0.778 | 0.779 | 0.780 |
| | 300 | 0.778 | 0.001 | 0.775 | 0.777 | 0.778 | 0.779 | 0.780 |
| | 400 | 0.778 | 0.001 | 0.775 | 0.777 | 0.778 | 0.779 | 0.780 |
| | 500 | 0.778 | 0.001 | 0.775 | 0.777 | 0.778 | 0.779 | 0.780 |
| | 1000 | 0.778 | 0.001 | 0.775 | 0.777 | 0.778 | 0.779 | 0.780 |
| | 1500 | 0.778 | 0.001 | 0.775 | 0.777 | 0.778 | 0.779 | 0.780 |
| | 2000 | 0.778 | 0.001 | 0.775 | 0.777 | 0.778 | 0.779 | 0.780 |
| 2000 | 100 | 0.778 | 0.001 | 0.775 | 0.777 | 0.778 | 0.779 | 0.780 |
| | 200 | 0.778 | 0.001 | 0.775 | 0.777 | 0.778 | 0.779 | 0.780 |
| | 300 | 0.778 | 0.001 | 0.775 | 0.777 | 0.778 | 0.779 | 0.780 |
| | 400 | 0.778 | 0.001 | 0.775 | 0.777 | 0.778 | 0.779 | 0.780 |
| | 500 | 0.778 | 0.001 | 0.775 | 0.777 | 0.778 | 0.779 | 0.780 |
| | 1000 | 0.778 | 0.001 | 0.775 | 0.777 | 0.778 | 0.779 | 0.780 |
| | 1500 | 0.778 | 0.001 | 0.775 | 0.777 | 0.778 | 0.779 | 0.780 |
| | 2000 | 0.778 | 0.001 | 0.775 | 0.777 | 0.778 | 0.779 | 0.780 |

**e-Table 4.** Distribution of the upper 95% confidence limit of the AUC for the GUSTO-I dataset using the 8-predictor model with the two-stage bootstrap confidence interval through 5000 replications: the 0.632 method.

| $B_{outer}$ | $B_{inner}$ | Mean | SD | Quantiles | | | | |
|---|---|---|---|---|---|---|---|---|
| | | | | 2.5% | 25% | 50% | 75% | 97.5% |
| 500 | 100 | 0.850 | 0.002 | 0.846 | 0.848 | 0.849 | 0.851 | 0.854 |
| | 200 | 0.849 | 0.002 | 0.846 | 0.848 | 0.849 | 0.851 | 0.854 |
| | 300 | 0.849 | 0.002 | 0.846 | 0.848 | 0.849 | 0.851 | 0.853 |
| | 400 | 0.849 | 0.002 | 0.846 | 0.848 | 0.849 | 0.851 | 0.853 |
| | 500 | 0.849 | 0.002 | 0.846 | 0.848 | 0.849 | 0.851 | 0.853 |
| | 1000 | 0.849 | 0.002 | 0.846 | 0.848 | 0.849 | 0.851 | 0.853 |
| | 1500 | 0.849 | 0.002 | 0.846 | 0.848 | 0.849 | 0.851 | 0.853 |
| | 2000 | 0.849 | 0.002 | 0.846 | 0.848 | 0.849 | 0.851 | 0.853 |
| 1000 | 100 | 0.850 | 0.001 | 0.847 | 0.849 | 0.850 | 0.851 | 0.852 |
| | 200 | 0.850 | 0.001 | 0.847 | 0.849 | 0.850 | 0.851 | 0.852 |
| | 300 | 0.850 | 0.001 | 0.847 | 0.849 | 0.850 | 0.850 | 0.852 |
| | 400 | 0.850 | 0.001 | 0.847 | 0.849 | 0.850 | 0.850 | 0.852 |
| | 500 | 0.850 | 0.001 | 0.847 | 0.849 | 0.850 | 0.850 | 0.852 |
| | 1000 | 0.850 | 0.001 | 0.847 | 0.849 | 0.850 | 0.850 | 0.852 |
| | 1500 | 0.850 | 0.001 | 0.847 | 0.849 | 0.850 | 0.850 | 0.852 |
| | 2000 | 0.850 | 0.001 | 0.847 | 0.849 | 0.850 | 0.850 | 0.852 |
| 1500 | 100 | 0.850 | 0.001 | 0.847 | 0.849 | 0.850 | 0.850 | 0.852 |
| | 200 | 0.850 | 0.001 | 0.847 | 0.849 | 0.850 | 0.850 | 0.852 |
| | 300 | 0.850 | 0.001 | 0.847 | 0.849 | 0.850 | 0.850 | 0.852 |
| | 400 | 0.850 | 0.001 | 0.847 | 0.849 | 0.850 | 0.850 | 0.852 |
| | 500 | 0.850 | 0.001 | 0.847 | 0.849 | 0.850 | 0.850 | 0.852 |
| | 1000 | 0.850 | 0.001 | 0.847 | 0.849 | 0.850 | 0.850 | 0.852 |
| | 1500 | 0.850 | 0.001 | 0.847 | 0.849 | 0.850 | 0.850 | 0.852 |
| | 2000 | 0.850 | 0.001 | 0.847 | 0.849 | 0.850 | 0.850 | 0.852 |
| 2000 | 100 | 0.850 | 0.001 | 0.848 | 0.849 | 0.850 | 0.850 | 0.852 |
| | 200 | 0.850 | 0.001 | 0.848 | 0.849 | 0.850 | 0.850 | 0.852 |
| | 300 | 0.850 | 0.001 | 0.848 | 0.849 | 0.850 | 0.850 | 0.852 |
| | 400 | 0.850 | 0.001 | 0.848 | 0.849 | 0.850 | 0.850 | 0.852 |
| | 500 | 0.850 | 0.001 | 0.848 | 0.849 | 0.850 | 0.850 | 0.852 |
| | 1000 | 0.850 | 0.001 | 0.848 | 0.849 | 0.850 | 0.850 | 0.852 |
| | 1500 | 0.850 | 0.001 | 0.848 | 0.849 | 0.850 | 0.850 | 0.852 |
| | 2000 | 0.850 | 0.001 | 0.848 | 0.849 | 0.850 | 0.850 | 0.852 |

**e-Table 5.** Distribution of the lower 95% confidence limit of the AUC for the GUSTO-I dataset using the 8-predictor model with the two-stage bootstrap confidence interval through 5000 replications: the 0.632+ method.

| $B_{outer}$ | $B_{inner}$ | Mean | SD | Quantiles | | | | |
| --- | --- | --- | --- | --- | --- | --- | --- | --- |
| | | | | 2.5% | 25% | 50% | 75% | 97.5% |
| 500 | 100 | 0.778 | 0.002 | 0.773 | 0.776 | 0.778 | 0.779 | 0.782 |
| | 200 | 0.778 | 0.002 | 0.773 | 0.776 | 0.778 | 0.779 | 0.782 |
| | 300 | 0.778 | 0.002 | 0.773 | 0.776 | 0.778 | 0.779 | 0.782 |
| | 400 | 0.778 | 0.002 | 0.773 | 0.776 | 0.778 | 0.779 | 0.782 |
| | 500 | 0.778 | 0.002 | 0.773 | 0.776 | 0.778 | 0.779 | 0.782 |
| | 1000 | 0.778 | 0.002 | 0.773 | 0.776 | 0.778 | 0.780 | 0.782 |
| | 1500 | 0.778 | 0.002 | 0.773 | 0.776 | 0.778 | 0.780 | 0.782 |
| | 2000 | 0.778 | 0.002 | 0.773 | 0.776 | 0.778 | 0.780 | 0.782 |
| 1000 | 100 | 0.778 | 0.002 | 0.774 | 0.776 | 0.778 | 0.779 | 0.781 |
| | 200 | 0.778 | 0.002 | 0.774 | 0.777 | 0.778 | 0.779 | 0.781 |
| | 300 | 0.778 | 0.002 | 0.774 | 0.777 | 0.778 | 0.779 | 0.781 |
| | 400 | 0.778 | 0.002 | 0.774 | 0.777 | 0.778 | 0.779 | 0.781 |
| | 500 | 0.778 | 0.002 | 0.774 | 0.777 | 0.778 | 0.779 | 0.781 |
| | 1000 | 0.778 | 0.002 | 0.774 | 0.777 | 0.778 | 0.779 | 0.781 |
| | 1500 | 0.778 | 0.002 | 0.774 | 0.777 | 0.778 | 0.779 | 0.781 |
| | 2000 | 0.778 | 0.002 | 0.774 | 0.777 | 0.778 | 0.779 | 0.781 |
| 1500 | 100 | 0.777 | 0.001 | 0.775 | 0.777 | 0.778 | 0.778 | 0.780 |
| | 200 | 0.778 | 0.001 | 0.775 | 0.777 | 0.778 | 0.779 | 0.780 |
| | 300 | 0.778 | 0.001 | 0.775 | 0.777 | 0.778 | 0.779 | 0.780 |
| | 400 | 0.778 | 0.001 | 0.775 | 0.777 | 0.778 | 0.779 | 0.780 |
| | 500 | 0.778 | 0.001 | 0.775 | 0.777 | 0.778 | 0.779 | 0.780 |
| | 1000 | 0.778 | 0.001 | 0.775 | 0.777 | 0.778 | 0.779 | 0.780 |
| | 1500 | 0.778 | 0.001 | 0.775 | 0.777 | 0.778 | 0.779 | 0.780 |
| | 2000 | 0.778 | 0.001 | 0.775 | 0.777 | 0.778 | 0.779 | 0.780 |
| 2000 | 100 | 0.777 | 0.001 | 0.775 | 0.777 | 0.777 | 0.778 | 0.780 |
| | 200 | 0.778 | 0.001 | 0.775 | 0.777 | 0.778 | 0.778 | 0.780 |
| | 300 | 0.778 | 0.001 | 0.775 | 0.777 | 0.778 | 0.778 | 0.780 |
| | 400 | 0.778 | 0.001 | 0.775 | 0.777 | 0.778 | 0.778 | 0.780 |
| | 500 | 0.778 | 0.001 | 0.775 | 0.777 | 0.778 | 0.778 | 0.780 |
| | 1000 | 0.778 | 0.001 | 0.775 | 0.777 | 0.778 | 0.778 | 0.780 |
| | 1500 | 0.778 | 0.001 | 0.775 | 0.777 | 0.778 | 0.778 | 0.780 |
| | 2000 | 0.778 | 0.001 | 0.775 | 0.777 | 0.778 | 0.778 | 0.780 |

e-Table 6. Distribution of the upper 95% confidence limit of the AUC for the GUSTO-I dataset using the 8-predictor model with the two-stage bootstrap confidence interval through 5000 replications: the 0.632+ method.

| $B_{outer}$ | $B_{inner}$ | Mean | SD | Quantiles | | | | |
|---|---|---|---|---|---|---|---|---|
| | | | | 2.5% | 25% | 50% | 75% | 97.5% |
| 500 | 100 | 0.849 | 0.002 | 0.846 | 0.848 | 0.849 | 0.851 | 0.853 |
| | 200 | 0.849 | 0.002 | 0.846 | 0.848 | 0.849 | 0.851 | 0.853 |
| | 300 | 0.849 | 0.002 | 0.846 | 0.848 | 0.849 | 0.851 | 0.853 |
| | 400 | 0.849 | 0.002 | 0.846 | 0.848 | 0.849 | 0.851 | 0.853 |
| | 500 | 0.849 | 0.002 | 0.846 | 0.848 | 0.849 | 0.851 | 0.853 |
| | 1000 | 0.849 | 0.002 | 0.846 | 0.848 | 0.849 | 0.851 | 0.853 |
| | 1500 | 0.849 | 0.002 | 0.846 | 0.848 | 0.849 | 0.851 | 0.853 |
| | 2000 | 0.849 | 0.002 | 0.846 | 0.848 | 0.849 | 0.851 | 0.853 |
| 1000 | 100 | 0.850 | 0.001 | 0.847 | 0.849 | 0.850 | 0.850 | 0.852 |
| | 200 | 0.849 | 0.001 | 0.847 | 0.849 | 0.849 | 0.850 | 0.852 |
| | 300 | 0.849 | 0.001 | 0.847 | 0.849 | 0.849 | 0.850 | 0.852 |
| | 400 | 0.849 | 0.001 | 0.847 | 0.849 | 0.849 | 0.850 | 0.852 |
| | 500 | 0.849 | 0.001 | 0.847 | 0.849 | 0.849 | 0.850 | 0.852 |
| | 1000 | 0.849 | 0.001 | 0.847 | 0.849 | 0.849 | 0.850 | 0.852 |
| | 1500 | 0.849 | 0.001 | 0.847 | 0.848 | 0.849 | 0.850 | 0.852 |
| | 2000 | 0.849 | 0.001 | 0.847 | 0.848 | 0.849 | 0.850 | 0.852 |
| 1500 | 100 | 0.850 | 0.001 | 0.847 | 0.849 | 0.850 | 0.850 | 0.852 |
| | 200 | 0.850 | 0.001 | 0.847 | 0.849 | 0.850 | 0.850 | 0.852 |
| | 300 | 0.850 | 0.001 | 0.847 | 0.849 | 0.850 | 0.850 | 0.852 |
| | 400 | 0.850 | 0.001 | 0.847 | 0.849 | 0.850 | 0.850 | 0.852 |
| | 500 | 0.850 | 0.001 | 0.847 | 0.849 | 0.850 | 0.850 | 0.852 |
| | 1000 | 0.850 | 0.001 | 0.847 | 0.849 | 0.850 | 0.850 | 0.852 |
| | 1500 | 0.850 | 0.001 | 0.847 | 0.849 | 0.849 | 0.850 | 0.852 |
| | 2000 | 0.850 | 0.001 | 0.847 | 0.849 | 0.850 | 0.850 | 0.852 |
| 2000 | 100 | 0.850 | 0.001 | 0.848 | 0.849 | 0.850 | 0.850 | 0.852 |
| | 200 | 0.850 | 0.001 | 0.848 | 0.849 | 0.850 | 0.850 | 0.852 |
| | 300 | 0.850 | 0.001 | 0.848 | 0.849 | 0.850 | 0.850 | 0.852 |
| | 400 | 0.850 | 0.001 | 0.848 | 0.849 | 0.850 | 0.850 | 0.852 |
| | 500 | 0.850 | 0.001 | 0.848 | 0.849 | 0.850 | 0.850 | 0.852 |
| | 1000 | 0.850 | 0.001 | 0.848 | 0.849 | 0.850 | 0.850 | 0.851 |
| | 1500 | 0.850 | 0.001 | 0.848 | 0.849 | 0.850 | 0.850 | 0.851 |
| | 2000 | 0.850 | 0.001 | 0.848 | 0.849 | 0.850 | 0.850 | 0.851 |

## e-Appendix C: Additional simulation studies

We conducted additional simulations that (1) varied the expected event fractions to 0.25 and 0.50, and (2) used resamples of the covariates from the original GUSTO-I dataset. The other settings were the same as those in the simulation studies in Section 4. The results are presented as nested-loop plots in e-Figures 1 and 2, respectively.

When using expected event fractions of 0.25 and 0.50, the proposed confidence intervals showed favourable coverage performances that were similar to those in the main simulations (in Section 5). Note that the total sample sizes of these simulations are relatively small compared with the main simulations, because the sample size $n$ is determined by (the number of candidate predictor variables × EPV) / (expected event fraction). Although the coverage rates of the location-shifted method were relatively small compared with the main simulations, the only reason is the sample size reduction. However, the coverage properties of the location-shifted confidence intervals were clearly better than those of the conventional methods. Also, the two-stage bootstrap confidence intervals showed consistently favourable coverage properties.

When assessing performances using resamples of the GUSTO-I dataset covariates, the correlation structures of the covariate distributions of the GUSTO-I dataset were completely replicated. The overall results were similar to those in the main simulations (in Section 5), and the correlations did not strongly influence the overall performances.

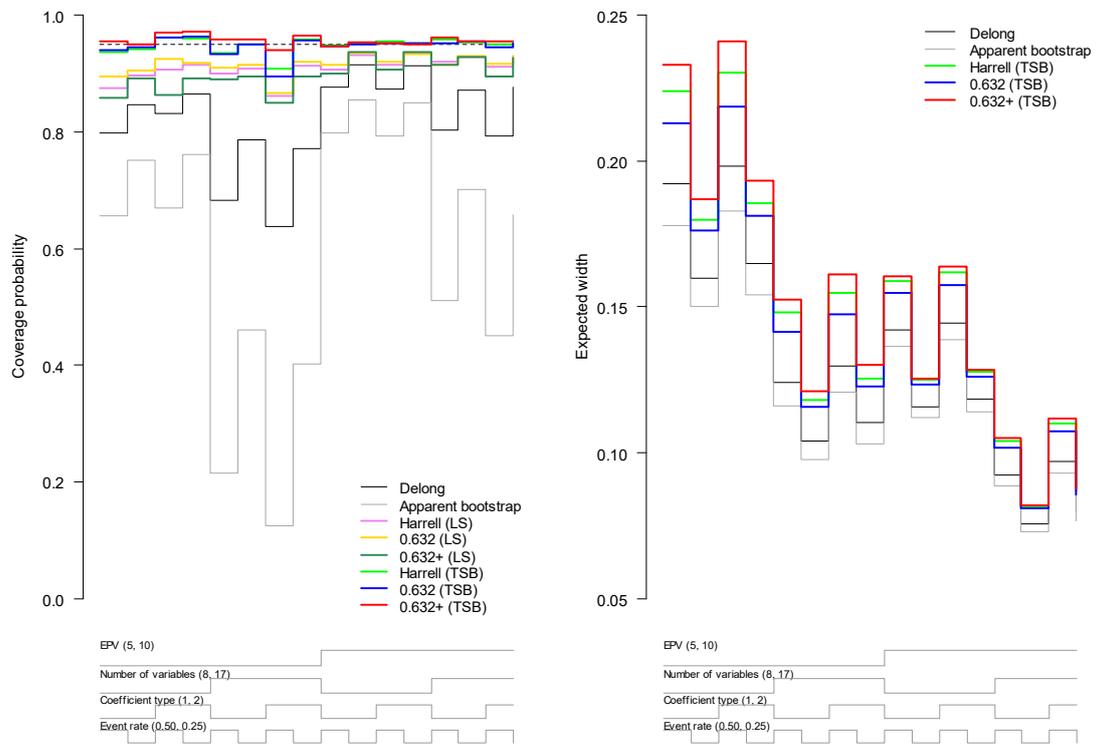

**e-Figure 1.** The coverage rates and expected widths of 95% confidence intervals in the simulation studies by the DeLong method, the apparent bootstrap confidence interval, the location-shifted (LS) confidence intervals, and the two-stage bootstrap (TSB) confidence intervals for the Harrell, 0.632, and 0.632+ methods.

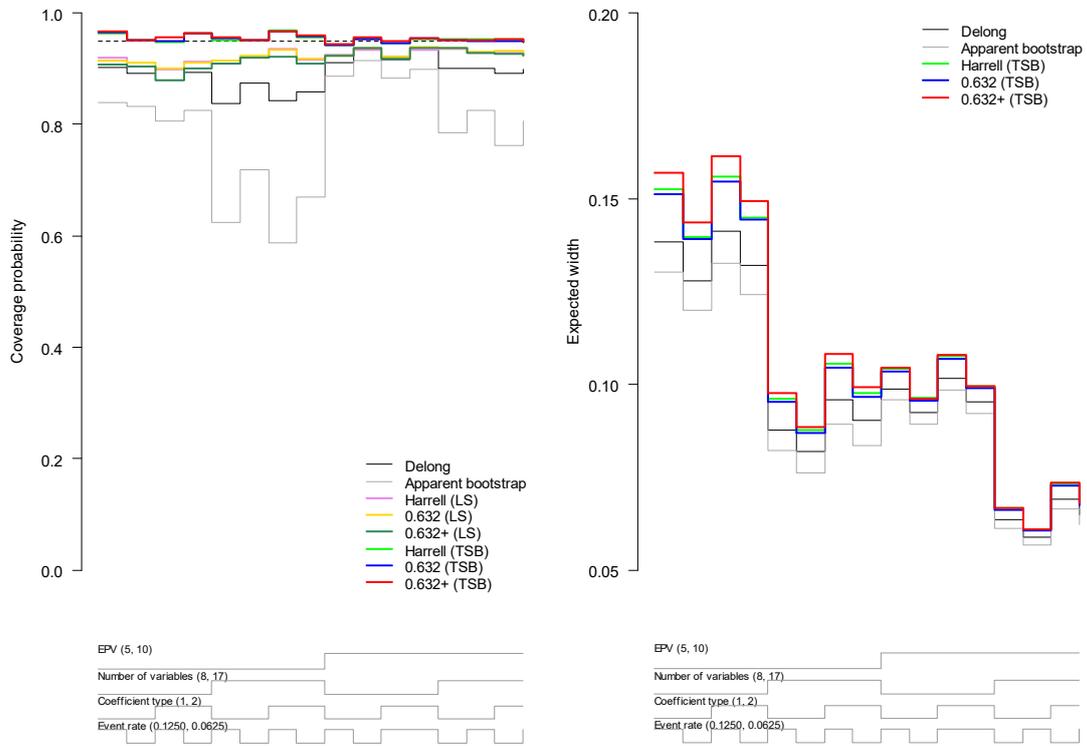

**e-Figure 2.** The coverage rates and expected widths of 95% confidence intervals in the simulation studies by the DeLong method, the apparent bootstrap confidence interval, the location-shifted (LS) confidence intervals, and the two-stage bootstrap (TSB) confidence intervals for the Harrell, 0.632, and 0.632+ methods.